\documentclass[prd,aps,preprintnumbers,fleqn,showpacs,nofootinbib,superscriptaddress]{revtex4}

\usepackage[utf8]{inputenc}
\usepackage[T1]{fontenc}
\usepackage{slashed}
\usepackage{times}

\usepackage{amssymb,amsfonts,amsmath,amsthm}
\usepackage{dsfont,bbm}
\usepackage{dcolumn}

\usepackage{graphicx}
\usepackage[caption=false]{subfig}

\begin{document}

\title{Sea quark Sivers distribution}

\author{Hui~Dong}
\affiliation{School of Physics and Key Laboratory of Particle Physics and Particle Irradiation
(MOE),  Shandong University, QingDao, Shandong 266237, China}
\author{Du-xin~Zheng}
\affiliation{School of Physics and Key Laboratory of Particle Physics and Particle Irradiation
(MOE),  Shandong University,  QingDao, Shandong 266237, China}
\author{Jian~Zhou} \affiliation{School of Physics and Key Laboratory of Particle Physics and Particle Irradiation
(MOE),  Shandong University,  QingDao, Shandong 266237, China}

\begin{abstract}
\noindent  We compute sea quark Sivers distribution within color
glass condensate(CGC) framework.  It has been found that up to the
leading logarithm accuracy,  the collinear twist-3 approach and the
CGC calculation yield the same result for sea quark Sivers
distribution in the dilute limit.  We further verify that transverse
momentum dependent factorization is consistent with CGC treatment at
small $x$ for the case of transverse spin asymmetry in open charm
quark production in semi-inclusive deeply inelastic scattering
process in an overlap kinematical region.
\end{abstract}

 \maketitle

\section{Introduction }

Polarization dependent phenomenology at small $x$ has attracted a lot of attentions in recent years, as it plays
an important role in studying gluon tomography of nucleon/nuclei. Employing powerful
color glass condensate(CGC) effective theory~\cite{McLerran:1993ni}
 and transverse momentum dependent(TMD) factorization~\cite{Collins:1981uk,collins} to polarization
effects at small $x$ has produced fruitful results. Among various topical issues in small $x$ spin physics,
 small $x$ parton helicity distributions~\cite{Bartels:1995iu,Kovchegov:2015pbl},
 linearly polarization of small $x$
 gluons~\cite{Metz:2011wb,Dominguez:2011br,Schafer:2012yx,Akcakaya:2012si,Dumitru:2015gaa,Boer:2017xpy,Benic:2017znu,Lansberg:2017dzg,Marquet:2017xwy},
(for a recent review covering this topic, see~\cite{Petreska:2018cbf}),
spin independent odderon~\cite{Kovchegov:2012ga,Hatta:2016aoc,Boer:2018vdi} and spin dependent odderon~\cite{Zhou:2013gsa,Boer:2015pni,Szymanowski:2016mbq},
 and elliptical gluon distribution~\cite{Hatta:2016dxp,Zhou:2016rnt,Hagiwara:2017ofm,Hatta:2017cte,Hagiwara:2017fye}
  were most extensively investigated recently. In this paper,
 we further explore the phenomenological consequence of spin dependent odderon.

In perturbative QCD,  odderon is a color-singlet exchange and can be formed by three gluons in a
symmetric color state. In the saturation regime, the expectation value of  odderon that incorporates multiple gluon exchange
effect has been computed in the MV model~\cite{McLerran:1993ni} with a cubic term~\cite{Jeon:2004rk}.
The energy dependence of the odderon exchange is described by the BKP
equation~\cite{Bartels:1980pe}, which also can be formulated in the dipole model~\cite{Kovchegov:2003dm}
and CGC framework~\cite{Hatta:2005as,Lappi:2016gqe}.
It was found in Ref.~\cite{Zhou:2013gsa} that a spin dependent odderon can be induced by an asymmetric color source distribution
(i.e. valence quark distribution in the context of the MV model)
in the transverse plane of a transversely polarized nucleon~\cite{Burkardt:2000za,Burkardt:2002ks,Gockeler:2006zu}.
To be more precise, transverse momentum transferred through spin dependent odderon is correlated with transverse
spin of target. In a more recent work~\cite{Boer:2015pni},
 three T-odd gluon TMDs are shown to be identical and related to the spin dependent odderon.
 It turns out that the spin dependent odderon is the only possible source contributing to transverse single spin asymmetries(SSAs) at
 small $x$ in the context of TMD factorization and CGC framework.

During the past few decades, transverse single spin asymmetries in high energy scattering is
one of the major focus in hadron physics studies.  It not only poses the great
theoretical challenge to account for the observed large SSAs, but also offers us  opportunities to address
some central aspects of hadron physics, such as the universality issue associated with QCD factorization theorem,
 and parton orbital angular momentum inside nucleon. Depending on kinematical regions, SSAs can be described
  within TMD factorization formalism~\cite{Sivers:1990fh,Collins:1992kk},
  or collinear twist-3 factorization approach~\cite{Efremov:1981sh,Qiu:1991pp,Yuan:2009dw}.
At small $x$, gluon Sivers  function
  and sea quark Sivers function dominate SSAs in the context of TMD factorization
  (for the relevant phenomenology work, see Refs.~\cite{Boer:2016fqd,Lu:2016vqu,Mukherjee:2016qxa,DAlesio:2017rzj,Boer:2015vso}), while
 SSAs at small $x$ is generated by the C-odd tri-gluon correlation function
 in the collinear twist-3 factorization~\cite{Koike:2011mb,Beppu:2010qn}.
 Gluon Sivers function is one of three T-odd gluon TMDs that
are proportional to the spin dependent odderon. In addition, it was shown that the $k_\perp$ moment of the
spin dependent odderon(or gluon Sivers function) can be related to the C-odd tri-gluon correlation function~\cite{Zhou:2013gsa}.

In the present work, we argue that sea quark Sivers distribution can be dynamically generated through the spin
dependent odderon.  To be more specific, in the semi-hard region, one can perturbatively compute sea quark Sivers distribution in CGC
formalism, and express it as the convolution of hard coefficient and the dipole amplitude.
 Due to the T-odd nature of the Sivers function, only the imaginary part of the dipole amplitude
that is identified as odderon contributes to the sea quark Sivers distribution. It is easy to verify that the resulting
sea quark Sivers function is consistent with that computed in the collinear twist-3 approach in the dilute limit
up to the leading logarithm accuracy.  The SSA for open charm production
in semi-inclusive DIS process at small $x$ has been computed in CGC formalism in an earlier work~\cite{Zhou:2013gsa}. With the
derived sea quark Sivers function, we are able to recover TMD formalism result for this observable in the
kinematical region where the produced open charm transverse momentum is much smaller than the virtual photon
off shellness  $Q$. In view of these findings, it might be fair to claim that the spin dependent odderon plays a central role in
describing SSAs phenomenology in the small $x$ region.

The paper is structured as follows. In the next section, we present the calculation in details together with
some numerical results for sea quark Sivers distribution with TMD evolution effect being incorporated.
In the Sec.~III, we show how to reduce the CGC result for the SSA in SIDIS to that derived in
TMD factorization. A summary of our findings and conclusions is presented in Sec.~IV.

\section{Sea quark sivers distribution in CGC }
Since the number density of gluons rises very rapidly towards small $x$ region,
 sea quark production is dominated by gluon splitting channel. The spin independent sea quark
 distribution has been computed in the dipole model~\cite{Mueller:1999wm}
 and in the CGC formalism~\cite{McLerran:1998nk}. The calculation of
 sea quark Sivers function can be formulated in a similar manner.  The target polarization
  dependence enters  the formula through the spin dependent odderon.
We start our calculation with introducing the operator definition for the odderon.
 Within the CGC formalism, one can identify the following operator
as the dipole odderon operator~\cite{Hatta:2005as},
\begin{eqnarray}
\hat O(b_\perp,r_\perp)=\frac{1}{2i} \left [ \hat D(b_\perp,r_\perp)-\hat D(b_\perp, -r_\perp) \right ] \ ,
\end{eqnarray}
where
\begin{eqnarray}
\hat D(b_\perp, r_\perp)=\frac{1}{N_c}
{\rm Tr} \left [ U(b_\perp+\frac{r_\perp}{2}) U^\dag(b_\perp-\frac{r_\perp}{2} ) \right ] \ ,
\end{eqnarray}
with the Wilson line in the fundamental representation  being given by,
$U(x_\perp)= {\cal P} e^{ig \int_{-\infty}^{+\infty} dx^- A_+(x^-, \ x_\perp)}  $.
Here, the plus and minus light cone components are defined in a common way. Obviously,
 odderon operator changes sign when exchanging the transverse positions of two Wilson lines.
As a consequence odderon contribution is the $k_\perp$-odd part of the dipole amplitude
 in momentum space. Motivated by this
observation, one can alternatively identify odderon contribution
 by parameterizing dipole amplitude in the following way,
\begin{eqnarray}
&&\int \frac{d^2 b_\perp d^2r_\perp  }{{(2\pi)^2}} e^{-ik_{\perp}
\cdot r_\perp} \frac{1}{N_c}\langle U(b_\perp+\frac{r_\perp}{2})
U^\dagger(b_\perp-\frac{r_\perp}{2})\rangle_{x_g} \nonumber \\ &=&
F(x_g,k_{\perp}^2)+ \left [ k_\perp \cdot b_\perp \right ] O(x_g,k_\perp^2)+
\left [ \frac{1}{M}
 \epsilon_{\perp}^{ij}S_{\perp i}  k_{\perp j}  \right ] O_{1T}^\perp(x_g,k_{\perp}^2) +
 \left[ (k_\perp \cdot b_\perp)^2- \frac{1}{2}k_\perp^2 b_\perp^2  \right ]F^{\cal E}(x_g,k_\perp^2)
\end{eqnarray}
where the first term in the second line is the spin independent gluon distribution. The second  and
the third terms are spin independent odderon~\cite{Kovchegov:2012ga}
 and spin dependent odderon~\cite{Zhou:2013gsa} respectively.
$F^{\cal E}(x_g,k_\perp^2)$ is the so-called gluon elliptical distribution~\cite{Hatta:2016dxp}.
It is common practice to compute
various gluon distributions using the MV model in which valence quark inside nucleon/nuclei is treated as
color source. It has been found that odderon contribution vanishes if valence quark/color source is uniformly
distributed in the transverse plane of nucleon/nucleci. For the spin independent case, the odderon contribution
 arises from the transverse gradient of color source distribution~\cite{Kovchegov:2012ga}, while
 an axial asymmetrical valence quark distribution in the transverse plane of a transversely polarized
 nucleon gives rise to a spin dependent odderon~\cite{Zhou:2013gsa}.

In the following we compute small $x$ quark Sivers function  in terms of the two point function
$ U(b_\perp+\frac{r_\perp}{2}) U^\dagger(b_\perp-\frac{r_\perp}{2})$ and relate it to spin dependent odderon.
As well known, the Sivers function is process dependent.
 We thus specify the gauge link in the matrix element definition of  sea quark Sivers function
  to be the future
 pointing one which is built up through final state interaction in SIDIS process. There are three diagrams
 contributing to the quark production amplitude in CGC formalism. The calculation is rather straightforward.
 Here we only would like to mention one interesting feature that four point function
 shows up in the intermediate step of the calculation. However, after some algebraic manipulation introduced
 in Ref.~\cite{Xiao:2017yya}, four point function collapses into the two point one.
 The final expression can be cast into the following  form,
\begin{equation}
xf_1(x,l_\perp^2)+\frac{\epsilon_{\perp}^{ij}S_{\perp i}  l_{\perp
j}}{M}xf_{1T}^\perp(x,l_{\perp}^2)=\frac{N_c}{8\pi^4}\int d\xi\int
d^2k_{\perp}\left \{ F(x_g, k_{\perp}^2)+\frac{1}{M}
 \epsilon_{\perp}^{ij}S_{\perp i}  k_{\perp j} O_{1T}^\perp(x_g,k_{\perp}^2) \right \}
  A(l_{\perp},k_\perp) \label{sivers} \ ,
\end{equation}
where $\xi=x/x_g$ with $x_g$ standing for the momentum fraction of the incoming gluon. The
coefficient $A$ is given by,
\begin{equation}
A(l_{\perp},k_\perp)
=\left[\frac{l_\perp|l_\perp-k_{\perp}|}{(1-\xi)l_\perp^2+\xi(l_\perp-k_{\perp})^2}-
\frac{l_\perp-k_{\perp}}{|l_\perp-k_{\perp}|}\right]^{2} \ .
\label{mass}
\end{equation}
One can easily read off the Sivers function from the Eq.~\ref{sivers},
\begin{equation}
\frac{\epsilon_{\perp}^{ij}S_{\perp i}  l_{\perp
j}}{M} xf_{1T}^\perp(x,l_{\perp}^2)=\frac{N_c}{8\pi^4}\int d\xi\int
d^2k_{\perp} \frac{1}{M}
 \epsilon_{\perp}^{ij}S_{\perp i}  k_{\perp j}
O_{1T}^\perp(x_g,k_{\perp}^2)
  A(l_{\perp},k_\perp) \ , \label{8}
\end{equation}
or alternatively,
\begin{equation}
xf_{1T}^\perp(x,l_{\perp}^2)=\frac{N_c}{8\pi^4}\int d\xi\int
d^2k_{\perp} \frac{k_\perp \cdot l_\perp}{l_\perp^2}
O_{1T}^\perp(x_g,k_{\perp}^2)
  A(l_{\perp},k_\perp) \ .
\end{equation}
which will be used as the initial condition when evolving quark Sivers function to higher scale with
the Collins-Soper evolution equation.

We notice that sea quark Sivers function has been computed in the collinear twist-3 approach~\cite{Koike:2011mb,Beppu:2010qn} as well.
It would be interesting
to check whether the CGC calculation and the collinear twist-3 approach produce the same result in the dilute limit($l_\perp^2 \gg Q_s^2 $)
 where they both apply.
 To this end, one can  extrapolate the CGC result to the dilute limit
  by Taylor expanding $A(l_{\perp},k_\perp)$ in terms of the power of $|k_\perp|/|l_\perp|$.
\begin{equation}
A(l_{\perp},k_\perp)=A(l_{\perp},k_\perp=0)+\frac{\partial A(l_{\perp},k_\perp)}{\partial k_\perp^\alpha} k_\perp^\alpha
+\frac{1}{2}\frac{\partial^2 A(l_{\perp},k_\perp)}{\partial k_\perp^\alpha \partial k_\perp^\beta} k_\perp^\alpha k_\perp^\beta
+\frac{1}{6}\frac{\partial^3 A(l_{\perp},k_\perp)}{\partial k_\perp^\alpha \partial k_\perp^\beta \partial k_\perp^\gamma} k_\perp^\alpha k_\perp^\beta k_\perp^\gamma+...
\end{equation}
where the first two terms vanish due to the Ward identity, and the third term only contributes to the spin independent distribution.
The leading power contribution to the sea quark Sivers distribution comes from the fourth term. Inserting this
expansion into Eq.~\ref{sivers}, one obtains,
\begin{equation}
\frac{\epsilon_{\perp}^{ij}S_{\perp i}  l_{\perp
j}}{M} xf_{1T}^\perp(x,l_{\perp}^2)=\frac{N_c}{8\pi^4}\int d\xi
\frac{1}{6}\frac{\partial^3 A(l_{\perp},k_\perp)}{\partial k_\perp^\alpha \partial k_\perp^\beta \partial k_\perp^\gamma}
\int d^2k_{\perp} \frac{1}{M}
 k_\perp^\alpha k_\perp^\beta k_\perp^\gamma \epsilon_{\perp}^{ij}S_{\perp i}  k_{\perp j}
O_{1T}^\perp(x_g,k_{\perp}^2)
\end{equation}
The next step is to invoke the relation between  the $k_\perp$ moment of the spin dependent odderon and
the collinear C-odd tri-gluon correlation function first derived in~\cite{Zhou:2013gsa},
\begin{eqnarray}
 \int d^2 k_\perp k_\perp^\alpha k_\perp^\beta k_\perp^\gamma
\frac{1}{M} \epsilon_{\perp}^{ij}S_{\perp i}  k_{\perp j} O_{1T}^\perp(x_g,k_\perp^2) = \frac{-i
g^2  \pi^2}{2 N_c }O^{\alpha \beta \gamma}(x_g)
\end{eqnarray}
where the C-odd tri-gluon correlation function is defined as,
\begin{eqnarray}
\!\!\!\!\!\!\!\!\!
O^{\alpha \beta \gamma}(x_g) \equiv  O^{\alpha \beta \gamma}(x_1,x_2) =ig
\int \frac{dy_1^- dy_2^-}{2\pi} e^{ix_1p^+y_1^-}e^{i(x_2-x_1)p^+y_2^-}\frac{1}{p^+}
\langle PS_\perp| d^{bca} F_b^{\beta+}(0) F_c^{\gamma+}(y_1^-)F_a^{a+}(y_2^-)|PS_\perp \rangle
\nonumber \\
=\frac{-i}{4\pi} \left [O(x_1,x_2) g_\perp^{\alpha \beta} \epsilon^{\gamma n p S_\perp}
+O(x_2,x_2-x_1) g_\perp^{\beta \gamma} \epsilon^{\alpha npS_\perp}
+O(x_1,x_1-x_2)g_\perp^{\gamma \alpha} \epsilon^{\beta npS_\perp}
\right ]
\end{eqnarray}
with $x_g$ being the total momentum transfer carried by gluons, which can be conveniently chosen
to be $x_g\equiv {\rm Max} \{ x_1,x_2  \}$. Note that our convention for the tri-gluon correlation
function differs from that defined in Ref.~\cite{Koike:2011mb,Beppu:2010qn} by a factor $-8\pi M$.
 If the symmetric tensor $d^{abc}$ is
replaced with the antisymmetric one $f^{abc}$ in the above matrix element, one can correspondingly define the
N type  C-even tri-gluon correlation function.

In the small $x$ limit, one has
 $O(x_1,x_2) \approx O(x_2,x_2-x_1) \approx O(x_1,x_1-x_2)\equiv O(x_g)$
 in the leading logarithm approximation. We then proceed by expressing the Sivers distribution in terms of
 tri-gluon correlation function $O(x_g)$,
 \begin{equation}
\frac{1}{M} xf_{1T}^\perp(x,l_{\perp}^2)=\frac{N_c}{8\pi^4}\frac{1}{l_\perp^4}
 \int d\xi \ \frac{1}{6}16(3\xi-8\xi^2+6\xi^3)\frac{-
g^2 \pi^2}{2 N_c }\frac{3}{4 \pi}O(x_g)
\end{equation}
Within the leading logarithm approximation, $O(x_g)$ can be viewed as the function that is independent of $x_g$.
\footnote{According to the BLV solution~\cite{Bartels:1999yt} to the BKP equation, the intercept of the odderon happens to be zero.
As a consequence, $O(x_g)$ is strictly independent of $x_g$.} It is  straightforward to carry out the integration over
$\xi$. One ends up with,
\begin{eqnarray}
\frac{1}{M}x
f_{1T}^\perp(x,l_{\perp}^2)|_{x\rightarrow0}=-\frac{\alpha_s}{2\pi^2}\frac{1}{l_\perp^4}
\frac{1}{3}O(x_g)
\end{eqnarray}

On the other hand, sea quark Sivers function computed in the collinear twist-3 approach reads~\cite{Ma:2012xn,Dai:2014ala},
\begin{eqnarray}
\frac{1}{M}f_{1T}^\perp(x,l_{\perp}^2)=-\frac{\alpha_s}{2\pi^2}\frac{1}{l_\perp^4}
\int_{x} \frac{dx_g}{x_g^2} \ \frac{\xi^2+(1-\xi)^2}{2} \frac{1}{2}
\left [ O(x_g,x_g)+O(x_g,0) \right ]
\end{eqnarray}
where the contribution from the C-even tri-gluon correlation
function $N(x_g,x_g)$ and $ N(x_g,0)$ has been neglected. This is justified to do so because the $O$ type function
is enhanced by the power of $\frac{1}{x_g}$ when taking into account small $x$ evolution
 with respect to the $N$ type function in the small $x$ limit~\cite{Schafer:2013opa}.
By identifying $O(x_g)\equiv O(x_g,x_g)=O(x_g,0)$ and carrying out $\xi$ integration,
 we find that the collinear
 twist-3 result is in complete agreement with that we derived in CGC formalism.  However, it is worth to emphasize again
 that such equivalence only can be established up to the leading logarithm accuracy.

 Let's close discussions on the analytical calculations with one final remark.  Following the similar procedure,
one also can evaluate small $x$ anti-quark Sivers function,
 which turns out to be the same as the quark Sivers function in magnitude, but with a reversed sign.
  This is in sharp contrast to the spin independent case where  small $x$
quark and anti-quark distributions computed in CGC formalism are identical.
 The relative minus sign can be best understood by noticing the C-odd nature of the spin dependent odderon.
 It might be promising to determine sea quark Sivers function
 by fitting to the experimental data on the SSA in $W^\pm$ production in pp collisions~\cite{Adamczyk:2015gyk}.
The $\bar u$ and $\bar d$ Sivers function extracted from this observable~\cite{Tian:2017qwk} seems to be
 compatible with our result in terms of sign.

We are  now ready to present some numerical results for sea quark Sivers distribution
by taking into account TMD evolution effect.  At tree level, both the unpolarized gluon distribution
and the spin dependent odderon can be computed in the MV model. They are given by,
\begin{eqnarray}
 F(x_g, k_\perp^2)&=&\pi R_0^2 \int \frac{d^2r_\perp }{(2\pi)^2} e^{-i  k_\perp \cdot r_\perp}
e^{-\frac{1}{4} r_\perp^2 Q_{s}^2} \
\\
O_{1T}^\perp(x_g,k_\perp^2)&=&
\frac{ -c_0 \alpha_s^3 \left ( \kappa_p^u + \kappa_p^d   \right )  }{4 R_0^4} \left [
\frac{\partial}{\partial k_\perp^2 } \frac{\partial}{\partial k_\perp^i} \frac{\partial}{\partial k_{\perp i}}
F(x_g,k_\perp^2) \right ] \ .
\label{polarized}
\end{eqnarray}
which will be used  as  the initial conditions when implementing TMD evolution.
In the above expression,  $ \kappa_p^u $ and $ \kappa_p^d $
are the contributions from up and down quarks  to the anomalous magnetic moment
of  proton, respectively. The color factor is given by $c_0=\frac{(N_c^2-1)(N_c^2-4)}{4 N_c^3} $.
 To facilitate numerical estimation, we reexpress the transverse area of nucleon as~\cite{Mueller:1999wm},
\begin{eqnarray}
\pi R_0^2=\frac{  2\pi^2 \alpha_s  x_g G(x_g,\mu_0)}{N_cQ^2_s}
 \label{eqn:ungluonGBW}
\end{eqnarray}
where  $G(x_g,\mu_0)$ is the standard gluon PDF in a nucleon, for which we will employ the MSTW
2008 LO PDF set.
In the following numerical estimations, the initial scale
$\mu_0$ is chosen to be $\mu_0=\text{0.55 GeV}$. In an earlier work~\cite{Boer:2017xpy}, we determined
the initial scale as the saturation scale. However, in the current case, only if
one chooses a fixed initial scale which doesn't change with varying $x_g$,
the ratio between $O_{1T}^\perp(x_g,k_\perp^2)$ and $F(x_g, k_\perp^2)$ roughly behaves
as $x^{0.3}$ that is compatible with the predication from the BFKL and the BKP equations.   The saturation scale is further fixed using
the GBW model~\cite{GolecBiernat:1998js}
$ Q^2_s(x)=(1\ \text{GeV})^2 A^{1/3}
\left(\frac{x_0}{x}\right)^{0.3}~~~\text{ with}~~ x_0=3\times 10^{-4}$.
\begin{figure}[t]
\begin{center}
\includegraphics[width=11 cm]{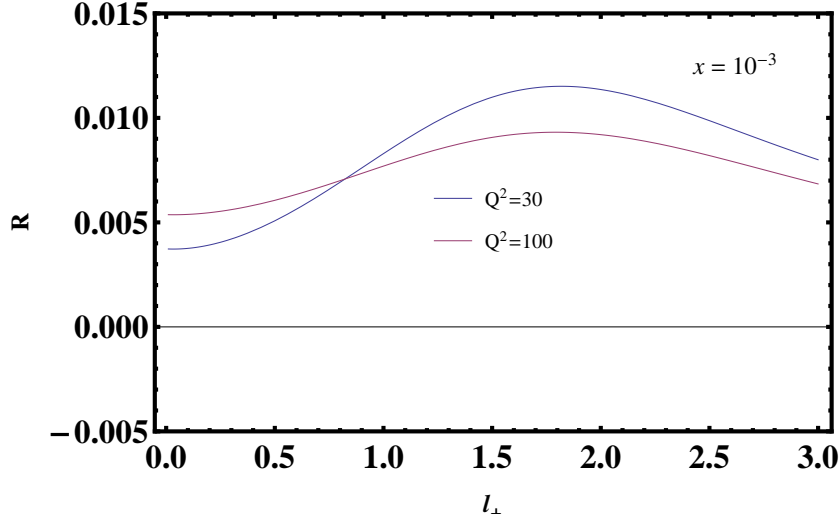}
\caption[] {The ratio R=$f_{1T}^\perp / f_1$ as the function of $l_\perp$, at $x=10^{-3}$ for
$Q^2=30 \text{GeV}^2$ and $Q^2=100 \text{GeV}^2$.}
 \label{1}
\end{center}
\end{figure}
\begin{figure}[t]
\begin{center}
\includegraphics[width=11 cm]{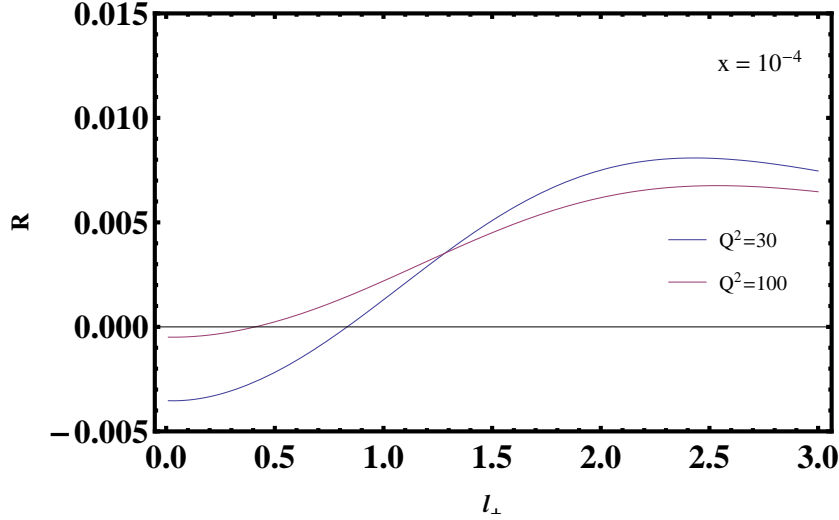}
\caption[] {The ratio R=$f_{1T}^\perp / f_1$ as the function of $l_\perp$, at $x=10^{-4}$ for
$Q^2=30 \text{GeV}^2$ and $Q^2=100 \text{GeV}^2$.}
 \label{1}
\end{center}
\end{figure}
The Collins-Soper evolution that governs the energy dependence of parton TMDs has been well
established in the moderate and large $x$ region~\cite{Collins:1981uk,collins}.
Recent progress~\cite{Mueller:2012uf,Balitsky:2015qba,Marzani:2015oyb,Zhou:2016tfe,Xiao:2017yya}
 suggests that small $x$ gluon TMDs
satisfy the same Collins-Soper equation. One would expect that the same analysis applies to the small $x$
quark TMD case. After solving the Collins-Soper equation for quark TMDs, all large logarithm terms
are resummed into an exponential, known as the Sudaokv factor. The evolved quark TMDs take form~\cite{collins,Ji:2004xq},
\begin{eqnarray}
 x f_{1} (x,l_{\perp}^2,\mu^2=Q^2)
 &=& \int d |b_{\perp}| ~|b_{\perp}| ~2 \pi J_0(|l_{\perp}|| b_{\perp}|)~
 e^{-S(\mu_b^2,Q^2)+S(\mu_b^2,\mu_0^2)}~
 x f_{1}(x,b_{\perp }^2,\mu_b^2), \\
  x f_{1T}^{\perp } (x,l_{\perp}^2,\mu^2=Q^2)&=&- \frac{1}{|l_\perp|}
  \int d |b_{\perp}| ~|b_{\perp}| ~2 \pi
  J_1(|l_{\perp}||b_{\perp}|)~ e^{-S(\mu_b^2,Q^2)+S(\mu_b^2,\mu_0^2)}~ x f_{1T}^{\perp' }(x,b_{\perp }^2,\mu_b^2),
\end{eqnarray}
where $\mu_b=2 e^{-\gamma_E}/|b_{\perp}| $,
the standard $\zeta$ parameter is chosen identical to the renormalization scale $\mu_b$ and
not shown here.  The unpolarized quark TMDs and the derivative of quark Sivers TMD at the initial scale $\mu_0$ in $b_\perp$
space are given by,
\begin{eqnarray}
  x f_{1}(x,b_{\perp}^2,\mu_0^2) &=& \int \frac{d^2 {l}_{\perp}}{(2 \pi)^2}
  ~e^{i {l}_{\perp}\cdot {b}_{\perp}}
 x f_{1}(x,l_{\perp}^2,\mu_0^2),
 \label{eqn:xf1DPk}
\\
 x f^{\perp '}_{1T}(x,b_{\perp}^2,\mu_0^2) &=& -\int \frac{d|l_{\perp}|}{2\pi} l_\perp^2
J_1(|b_{\perp}||l_{\perp}|) ~ x f^{\perp }_{1T}(x,l_{\perp}^2,\mu_0^2), \label{eqn:xh1DPk}
\end{eqnarray}
The standard treatment for the non-perturbative part applies to the
Sudakov factor $S(\mu_b^2,Q^2)$ which at one-loop order reads,
\begin{eqnarray}
 S(\mu_b^2,Q^2) =
\frac{C_F}{2\pi} \int^{Q^2}_{\mu_{b*}^2} \frac{d\mu^2}{\mu^2} \alpha_s(\mu) \left[ \ln
\frac{Q^2}{\mu^2} - \frac{3}{2} \right ]+S^{NP}(b_\perp^2,Q^2),
\end{eqnarray}
where $\mu_{b*}^2$ is defined as $\mu_{b*}^2=4e^{-2\gamma_E}/b_{\perp*}^2$, with $b_{\perp*}$
given by
$b_{\perp *}=\frac{b_{\perp}}{\sqrt{1+b_{\perp}^2/b^2_{\max}}} $.
 The parametrization for the non-perturbative Sudakov factor $S^{NP}(b_\perp^2,Q^2)$
 is taken  as~\cite{Aybat:2011zv},
\begin{equation}
S_{q}^{NP}(b_{\perp}^2,Q^2)=
\frac{1}{2}\left( g_1+g_2 \ln \frac{Q}{2 Q_0} + 2 g_1 g_3 \ln \frac{10 x x_0}{x_0+x} \right)
b_{\perp}^2,
\end{equation}
with $b_{\max}=1.5\, \text{GeV}^{-1},~ g_1=0.201\, \text{GeV}^2, ~ g_2=0.184\,
\text{GeV}^2,~g_3=-0.129,~x_0=0.009,~Q_0=1.6\, \text{GeV}$.
In our numerical estimation,  we used the one-loop running coupling constant $\alpha_s$, with
$n_f=3$ and $\Lambda_{\text{QCD}} = 216~ \text{MeV}$.

With these ingredients, we are able to reproduce all numerical results for unpolarized small $x$ quark TMDs
presented in Ref.~\cite{Marquet:2009ca}. We further evolve sea quark Sivers function from the initial
 scale where the MV model result is used as the initial input up to scales $Q^2=30 \ \text{GeV}^2$ and $Q^2=100\  \text{GeV}^2$.
As the spin asymmetry is determined by the ratio between unpolarized quark TMD and  quark Sivers TMD, we plot
the ratio as the function of $l_\perp$ at $x=10^{-3}$ and $x=10^{-4}$ in Fig.1 and Fig.2, respectively.
As expected, the ratio is suppressed with increasing energy and decreasing $x$.
One further observes that the ratios grows with increasing transverse momentum at low $|l_\perp|$, until it
 reaches a maximal value around $l_\perp=1.8 \text{GeV}$. The maximal value of the ratio is on the
 percentage level.
 It is worthy to mention that the absolute size of the Sivers function critically
 depends on the initial scale we used. However, the transverse momentum dependence and the energy dependence of
 the Sivers function we predicated is less model dependent.

\section{The SSA in open charm production in SIDIS: CGC V.S. TMD}
One of  observables that allows us to access small $x$ charm quark Sivers function
 is the SSA for open charm in SIDIS process.
The spin independent differential cross section for this process
has been calculated in the dipole model~\cite{Mueller:1999wm}
 and in the CGC formalism~\cite{McLerran:1998nk}.  It is straightforward to extend the calculation
 to the spin dependent case which reads~\cite{Zhou:2013gsa},
\begin{eqnarray}
\frac{d \sigma}{dx_B dz   dy  d^2 p_{h\perp}}&=&\frac{ \alpha_{em}^2 e_c^2}{2 \pi^3 y x_B}
\int_{z_h} \frac{dz}{z} \frac{D(z)}{z^2} \int \frac{d^2 k_{\perp}}{(2\pi)^2}
 H(k_\perp,l_\perp,Q^2)
\int \frac{d^2 x_\perp d^2y_\perp  }{{(2\pi)^2}} e^{-ik_{\perp} \cdot (x_\perp-y_\perp)} \langle
U(x_\perp) U^\dagger(y_\perp)\rangle_{x_g} \nonumber\\&=&  \frac{ \alpha_{em}^2 e_c^2N_c}{2 \pi^3
yx_B }\int_{z_h} \frac{dz}{z} \frac{D(z)}{z^2}
 \int d^2 k_{\perp} H(\hat \xi,k_\perp,l_\perp,Q^2)
\left [  F_{x_g}(k_\perp^2)+
 \frac{1}{M} \epsilon_{\perp}^{ ij}S_{\perp i}  k_{\perp j} O_{1T, x_g}^\perp(k_\perp^2) \right ] \ ,
 \label{17}
\end{eqnarray}
Here the common kinematical variables in SIDIS process
 are defined as $Q^2=-q \cdot q$, $x_B=Q^2/2P\cdot q$, $y=q\cdot P/P_e \cdot P$, $l_\perp=p_{h\perp}/z$
 and $\hat \xi=z_h/z=l \cdot P/P\cdot q$ where $l$, $P_e$, $P$, and $q$  are momenta for produced charm quark,
 incoming lepton and proton, and virtual photon, respectively.
The term proportional to the spin dependent odderon in the second line is responsible for the SSA for charm production.
 For simplicity, we  only taken into account the transverse polarized virtual photon
contribution to the differential cross section. The hard part $H(\hat \xi, k_\perp,l_\perp,Q^2)$ is given by,
\begin{eqnarray}
H(\hat \xi, k_\perp,l_\perp,Q^2)=\left [ 1-y+\frac{y^2}{2} \right ] \left [ \hat \xi^2+(1-\hat \xi)^2 \right
] \left [ \frac{l_\perp-k_\perp}{\rho+(l_\perp-k_\perp)^2}- \frac{l_\perp}{\rho+l_\perp^2} \right
]^2  \ .
\end{eqnarray}
where $\rho$ is defined as $\rho=\hat \xi(1-\hat \xi)Q^2$.

Extracting charm quark Sivers function in this process relies on the use of TMD factorization.
To establish TMD factorization, there must exist
  an additional large scale that is much larger than parton transverse momentum. In the current case,
TMD factorization can be applied in the
kinematical region where $Q^2$ is much larger than  $l_\perp^2$.
To recover TMD factorization formula, one has to
extrapolate the CGC result to this kinematical region by isolating the leading power contribution in terms of
$l_\perp^2/Q^2$. This analysis actually has been carried out for the unpolarized case in Ref.~\cite{Marquet:2009ca}.
 In the same manner,
 one can show that the CGC result reduces to that obtained in TMD factorization for the polarized case
as well.

One first notices that the integral in Eq.\ref{17} is dominated by the end point contribution
$\hat \xi\rightarrow 1$. We thus should make power expansion around this end point.  To this end,
we insert a delta function $\int d \xi \delta(\xi-1/(1+\Lambda^2/\rho))$ into Eq.\ref{17} with
$\Lambda^2=(1-\hat \xi) l_\perp^2+\hat \xi (l_\perp-k_\perp)^2$\cite{Marquet:2009ca}.
One proceeds by expanding this delta function,
\begin{equation}
\delta \left (\xi-\frac{1}{1+\Lambda^2/\rho} \right ) \approx \frac{1-\hat \xi}{ \xi}
\left[ \frac{ \delta ( 1-\hat \xi)}{1-\xi}+\frac{ \delta ( 1- \xi)}{1-\hat \xi} \right ]
\end{equation}
where the second delta function gives rise to power suppressed contribution.   We then can carry out
the integration over $\hat \xi$ using the first delta function, and obtain the spin dependent differential
cross section,
\begin{eqnarray}
\frac{d \Delta \sigma}{dx_B dz_h   dy  d^2 p_{h\perp}}= \frac{ \alpha_{em}^2 e_c^2N_c(1-y+\frac{y^2}{2} ) }{2 \pi^3 Q^2
yx_B }\frac{D(z_h)}{z_h^2}
 \int d \xi d^2 k_{\perp} \ A(k_\perp,l_\perp)
 \frac{1}{M} \epsilon_{\perp}^{ ij}S_{\perp i}  k_{\perp j} O_{1T}^\perp(x_g,k_\perp^2)  \ .
 \label{CGC}
\end{eqnarray}
On the other hand, the SSA also can be formulated in the TMD factorization framework,
\begin{eqnarray}
\frac{d \Delta \sigma}{dx_B   dy dz_h d^2 p_{h\perp}}=\frac{4 \pi \alpha_{em}^2 e_c^2
(1-y+\frac{y^2}{2} ) }{Q^2y}
 \int d^2 l_\perp d^2 p_\perp \frac{\epsilon_{\perp}^{ij}S_{\perp i}
l_{\perp j}}{M}f_{1T}^\perp(x_B, l_{\perp}^2) D(z_h, p_\perp^2) \delta^2(z_h
l_\perp+p_\perp-p_{h\perp}) \label{tmd}
\end{eqnarray}
To make contact with the results derived in the CGC formalism,
we ignore $p_\perp$ in the above formula. This approximation
is justified because  $p_\perp$  is typically of order of $\Lambda_{QCD}$, while $k_\perp$ is of order of
$Q_s$ which is much larger than $\Lambda_{QCD}$.
Using the expression for the Sivers function given in Eq.\ref{8},
 the spin dependent differential cross section Eq.\ref{CGC} is reduced to the TMD
 formula Eq.\ref{tmd} as expected.  In the kinematical region where $Q^2 \sim l_\perp^2 \gg k_\perp^2$, the SSA
 for open charm production can be calculated in collinear twist approach~\cite{Beppu:2010qn}.
 One would expect that
 the equivalence between the collinear twist-3 approach and CGC framework in describing this observable
 only can be achieved in the leading logarithm approximation as discussed in the previous section.

\section{Summary}

To summarize, we compute sea quark Sivers function in terms of the C-odd part of the dipole amplitude which is
identified as spin dependent odderon inside a transversely polarized target.
Due to the C-odd nature of odderon, the computed quark Sivers function and anti-quark Sivers function
are the same in size, but differ by a minus sign.
As a consistency check, we verified that in the dilute limit,
the CGC result for sea quark Sivers function is reduced to that obtained using  collinear twist-3 approach
 in the leading logarithm approximation.
We further show that sea quark Sivers function can be accessed through the SSA in SIDIS by justifying
TMD factorization formula from a full CGC calculation.   It is worth to mention again
that spin dependent odderon is related to three T-odd gluon TMDs inside a transversely polarized nucleon and
 the C-odd tri-gluon correlation~\cite{Zhou:2013gsa,Boer:2015pni}. In view of these findings and the fact
 that the spin dependent odderon is the dynamical origin of sea quark Sivers function,
 one may conclude that spin dependent odderon plays a central role in describing single spin asymmetries
phenomenology at small $x$.

\

\noindent {\it \bf Acknowledgments:} This work has been supported by  the National Science
Foundation of China under Grant No. 11675093 and 11475104.
 The partial support from the Major State Basic Research Development Program in China (No.
2014CB845406) is acknowledged.


\begin{thebibliography}{99}


\bibitem{McLerran:1993ni}
  L.~D.~McLerran and R.~Venugopalan,
  Phys.\ Rev.\ D {\bf 49}, 2233 (1994);
  Phys.\ Rev.\ D {\bf 49}, 3352 (1994).

\bibitem{Collins:1981uk}
  J.~C.~Collins and D.~E.~Soper,
  Nucl.\ Phys.\ B {\bf 193}, 381 (1981)
  Erratum: [Nucl.\ Phys.\ B {\bf 213}, 545 (1983)],
  doi:10.1016/0550-3213(81)90339-4;
  Nucl.\ Phys.\ B {\bf 194}, 445 (1982).
  doi:10.1016/0550-3213(82)90021-9.

   \bibitem{collins}
  J. C. Collins, {\it Foundations of perturbative QCD} (Cambridge  University Press, Cambridge,
  2011)


\bibitem{Bartels:1995iu}
  J.~Bartels, B.~I.~Ermolaev and M.~G.~Ryskin,
  Z.\ Phys.\ C {\bf 70}, 273 (1996)
  [hep-ph/9507271];
  Z.\ Phys.\ C {\bf 72}, 627 (1996)
  doi:10.1007/s002880050285, 10.1007/BF02909194
  [hep-ph/9603204].

\bibitem{Kovchegov:2015pbl}
  Y.~V.~Kovchegov, D.~Pitonyak and M.~D.~Sievert,
  JHEP {\bf 1601}, 072 (2016)
  Erratum: [JHEP {\bf 1610}, 148 (2016)]
  doi:10.1007/JHEP01(2016)072, 10.1007/JHEP10(2016)148
  [arXiv:1511.06737 [hep-ph]];
  Phys.\ Rev.\ Lett.\  {\bf 118}, no. 5, 052001 (2017)
  doi:10.1103/PhysRevLett.118.052001
  [arXiv:1610.06188 [hep-ph]];
  Phys.\ Rev.\ D {\bf 95}, no. 1, 014033 (2017)
  doi:10.1103/PhysRevD.95.014033
  [arXiv:1610.06197 [hep-ph]];
  Phys.\ Lett.\ B {\bf 772}, 136 (2017)
  doi:10.1016/j.physletb.2017.06.032
  [arXiv:1703.05809 [hep-ph]];
  JHEP {\bf 1710}, 198 (2017)
  doi:10.1007/JHEP10(2017)198
  [arXiv:1706.04236 [nucl-th]].

\bibitem{Metz:2011wb}
  A.~Metz and J.~Zhou,
  Phys.\ Rev.\ D {\bf 84}, 051503 (2011).

\bibitem{Dominguez:2011br}
  F.~Dominguez, J.~W.~Qiu, B.~W.~Xiao and F.~Yuan,
  Phys.\ Rev.\ D {\bf 85}, 045003 (2012)
  [arXiv:1109.6293 [hep-ph]].

\bibitem{Schafer:2012yx}
  A.~Sch\"afer and J.~Zhou,
  Phys.\ Rev.\ D {\bf 85}, 114004 (2012)
  [arXiv:1203.1534 [hep-ph]].

\bibitem{Akcakaya:2012si}
  E.~Akcakaya, A.~Sch\"afer and J.~Zhou,
  Phys.\ Rev.\ D {\bf 87}, no. 5, 054010 (2013)
  [arXiv:1208.4965 [hep-ph]].

\bibitem{Dumitru:2015gaa}
  A.~Dumitru, T.~Lappi and V.~Skokov,
  Phys.\ Rev.\ Lett.\  {\bf 115}, no. 25, 252301 (2015)
  [arXiv:1508.04438 [hep-ph]].

\bibitem{Boer:2017xpy}
  D.~Boer, P.~J.~Mulders, J.~Zhou and Y.~j.~Zhou,
  JHEP {\bf 1710}, 196 (2017)
  doi:10.1007/JHEP10(2017)196
  [arXiv:1702.08195 [hep-ph]].

\bibitem{Benic:2017znu}
  S.~Benic and A.~Dumitru,
  arXiv:1710.01991 [hep-ph].

\bibitem{Lansberg:2017dzg}
  J.~P.~Lansberg, C.~Pisano, F.~Scarpa and M.~Schlegel,
  arXiv:1710.01684 [hep-ph].

\bibitem{Marquet:2017xwy}
  C.~Marquet, C.~Roiesnel and P.~Taels,
  Phys.\ Rev.\ D {\bf 97}, no. 1, 014004 (2018)
  doi:10.1103/PhysRevD.97.014004
  [arXiv:1710.05698 [hep-ph]].

\bibitem{Petreska:2018cbf}
  E.~Petreska,
  arXiv:1804.04981 [hep-ph].

\bibitem{Kovchegov:2012ga}
  Y.~V.~Kovchegov and M.~D.~Sievert,
  Phys.\ Rev.\ D {\bf 86}, 034028 (2012)
  [Erratum-ibid.\ D {\bf 86}, 079906 (2012)].

\bibitem{Hatta:2016aoc}
  Y.~Hatta, Y.~Nakagawa, F.~Yuan, Y.~Zhao and B.~Xiao,
  Phys.\ Rev.\ D {\bf 95}, no. 11, 114032 (2017)
  doi:10.1103/PhysRevD.95.114032
  [arXiv:1612.02445 [hep-ph]].

\bibitem{Boer:2018vdi}
  D.~Boer, T.~van Daal, P.~J.~Mulders and E.~Petreska,
  arXiv:1805.05219 [hep-ph].

\bibitem{Zhou:2013gsa}
  J.~Zhou,
  Phys.\ Rev.\ D {\bf 89}, no. 7, 074050 (2014)
  doi:10.1103/PhysRevD.89.074050
  [arXiv:1308.5912 [hep-ph]].


\bibitem{Boer:2015pni}
  D.~Boer, M.~G.~Echevarria, P.~Mulders and J.~Zhou,
  Phys.\ Rev.\ Lett.\  {\bf 116}, no. 12, 122001 (2016)
  doi:10.1103/PhysRevLett.116.122001
  [arXiv:1511.03485 [hep-ph]].

\bibitem{Szymanowski:2016mbq}
  L.~Szymanowski and J.~Zhou,
  Phys.\ Lett.\ B {\bf 760}, 249 (2016)
  doi:10.1016/j.physletb.2016.06.055
  [arXiv:1604.03207 [hep-ph]].

\bibitem{Hatta:2016dxp}
  Y.~Hatta, B.~W.~Xiao and F.~Yuan,
  Phys.\ Rev.\ Lett.\  {\bf 116}, no. 20, 202301 (2016)
  doi:10.1103/PhysRevLett.116.202301
  [arXiv:1601.01585 [hep-ph]].

\bibitem{Zhou:2016rnt}
  J.~Zhou,
  Phys.\ Rev.\ D {\bf 94}, no. 11, 114017 (2016)
  doi:10.1103/PhysRevD.94.114017
  [arXiv:1611.02397 [hep-ph]].

\bibitem{Hagiwara:2017ofm}
  Y.~Hagiwara, Y.~Hatta, B.~W.~Xiao and F.~Yuan,
  Phys.\ Lett.\ B {\bf 771}, 374 (2017)
  doi:10.1016/j.physletb.2017.05.083
  [arXiv:1701.04254 [hep-ph]].

\bibitem{Hatta:2017cte}
  Y.~Hatta, B.~W.~Xiao and F.~Yuan,
  Phys.\ Rev.\ D {\bf 95}, no. 11, 114026 (2017)
  doi:10.1103/PhysRevD.95.114026
  [arXiv:1703.02085 [hep-ph]].

\bibitem{Hagiwara:2017fye}
  Y.~Hagiwara, Y.~Hatta, R.~Pasechnik, M.~Tasevsky and O.~Teryaev,
  Phys.\ Rev.\ D {\bf 96}, no. 3, 034009 (2017)  doi:10.1103/PhysRevD.96.034009
  [arXiv:1706.01765 [hep-ph]].

\bibitem{Jeon:2004rk}
  S.~Jeon and R.~Venugopalan,
  Phys.\ Rev.\ D {\bf 70}, 105012 (2004);
  Phys.\ Rev.\ D {\bf 71}, 125003 (2005).

\bibitem{Bartels:1980pe}
  J.~Bartels,
  Nucl.\ Phys.\ B {\bf 175}, 365 (1980).
  J.~Kwiecinski and M.~Praszalowicz,
  Phys.\ Lett.\ B {\bf 94}, 413 (1980).


\bibitem{Kovchegov:2003dm}
  Y.~V.~Kovchegov, L.~Szymanowski and S.~Wallon,
  Phys.\ Lett.\ B {\bf 586}, 267 (2004).

\bibitem{Hatta:2005as}
  Y.~Hatta, E.~Iancu, K.~Itakura and L.~McLerran,
  Nucl.\ Phys.\ A {\bf 760}, 172 (2005).

\bibitem{Lappi:2016gqe}
  T.~Lappi, A.~Ramnath, K.~Rummukainen and H.~Weigert,
  Phys.\ Rev.\ D {\bf 94}, no. 5, 054014 (2016)
  doi:10.1103/PhysRevD.94.054014
  [arXiv:1606.00551 [hep-ph]].


\bibitem{Burkardt:2000za}
  M.~Burkardt,
  Phys.\ Rev.\ D {\bf 62}, 071503 (2000)
  [Erratum-ibid.\ D {\bf 66}, 119903 (2002)];
  Int.\ J.\ Mod.\ Phys.\ A {\bf 18}, 173 (2003).

\bibitem{Burkardt:2002ks}
  M.~Burkardt,
  Phys.\ Rev.\ D {\bf 66}, 114005 (2002).


\bibitem{Gockeler:2006zu}
  M.~Gockeler {\it et al.}  [QCDSF and UKQCD Collaborations],
  Phys.\ Rev.\ Lett.\  {\bf 98}, 222001 (2007).


\bibitem{Sivers:1990fh}
  D.~W.~Sivers,
  Phys.\ Rev.\ D {\bf 43}, 261 (1991).

\bibitem{Collins:1992kk}
  J.~C.~Collins,
  Nucl.\ Phys.\ B {\bf 396}, 161 (1993).

\bibitem{Efremov:1981sh}
  A.~V.~Efremov and O.~V.~Teryaev,
  Sov.\ J.\ Nucl.\ Phys.\  {\bf 36}, 140 (1982)
  [Yad.\ Fiz.\  {\bf 36}, 242 (1982)];
  Phys.\ Lett.\ B {\bf 150}, 383 (1985).


\bibitem{Qiu:1991pp}
  J.-w.~Qiu and G.~F.~Sterman,
  Phys.\ Rev.\ Lett.\  {\bf 67}, 2264 (1991);
  Nucl.\ Phys.\ B {\bf 378}, 52 (1992).

\bibitem{Yuan:2009dw}
  F.~Yuan and J.~Zhou,
  Phys.\ Rev.\ Lett.\  {\bf 103}, 052001 (2009).

\bibitem{Boer:2016fqd}
  D.~Boer, P.~J.~Mulders, C.~Pisano and J.~Zhou,
  JHEP {\bf 1608}, 001 (2016)
  doi:10.1007/JHEP08(2016)001
  [arXiv:1605.07934 [hep-ph]].

\bibitem{Lu:2016vqu}
  Z.~Lu and B.~Q.~Ma,
  Phys.\ Rev.\ D {\bf 94}, no. 9, 094022 (2016)
  doi:10.1103/PhysRevD.94.094022
  [arXiv:1611.00125 [hep-ph]].

\bibitem{Mukherjee:2016qxa}
  A.~Mukherjee and S.~Rajesh,
  Eur.\ Phys.\ J.\ C {\bf 77}, no. 12, 854 (2017)
  doi:10.1140/epjc/s10052-017-5406-4
  [arXiv:1609.05596 [hep-ph]];
  S.~Rajesh, R.~Kishore and A.~Mukherjee,
  arXiv:1802.10359 [hep-ph].

\bibitem{DAlesio:2017rzj}
  U.~D'Alesio, F.~Murgia, C.~Pisano and P.~Taels,
  Phys.\ Rev.\ D {\bf 96}, no. 3, 036011 (2017)
  doi:10.1103/PhysRevD.96.036011
  [arXiv:1705.04169 [hep-ph]].

\bibitem{Boer:2015vso}
  D.~Boer, C.~Lorce, C.~Pisano and J.~Zhou,
  Adv.\ High Energy Phys.\  {\bf 2015}, 371396 (2015)
  doi:10.1155/2015/371396
  [arXiv:1504.04332 [hep-ph]].

\bibitem{Koike:2011mb}
  Y.~Koike and S.~Yoshida,
  Phys.\ Rev.\ D {\bf 84}, 014026 (2011).


\bibitem{Beppu:2010qn}
  H.~Beppu, Y.~Koike, K.~Tanaka and S.~Yoshida,
  Phys.\ Rev.\ D {\bf 82}, 054005 (2010).

\bibitem{Mueller:1999wm}
  A.~H.~Mueller,
  Nucl.\ Phys.\ B {\bf 558}, 285 (1999).

\bibitem{McLerran:1998nk}
  L.~D.~McLerran and R.~Venugopalan,
  Phys.\ Rev.\ D {\bf 59}, 094002 (1999).



\bibitem{Xiao:2017yya}
  B.~W.~Xiao, F.~Yuan and J.~Zhou,
  Nucl.\ Phys.\ B {\bf 921}, 104 (2017)
  doi:10.1016/j.nuclphysb.2017.05.012
  [arXiv:1703.06163 [hep-ph]].


\bibitem{Ma:2012xn}
  J.~P.~Ma and Q.~Wang,
  Phys.\ Lett.\ B {\bf 715}, 157 (2012)
  doi:10.1016/j.physletb.2012.07.036
  [arXiv:1205.0611 [hep-ph]].

\bibitem{Dai:2014ala}
  L.~Y.~Dai, Z.~B.~Kang, A.~Prokudin and I.~Vitev,
  Phys.\ Rev.\ D {\bf 92}, no. 11, 114024 (2015)
  doi:10.1103/PhysRevD.92.114024
  [arXiv:1409.5851 [hep-ph]].


\bibitem{Schafer:2013opa}
  A.~Schafer and J.~Zhou,
  arXiv:1308.4961 [hep-ph].


\bibitem{Bartels:1999yt}
  J.~Bartels, L.~N.~Lipatov and G.~P.~Vacca,
  Phys.\ Lett.\ B {\bf 477}, 178 (2000).

\bibitem{Adamczyk:2015gyk}
  L.~Adamczyk {\it et al.} [STAR Collaboration],
  Phys.\ Rev.\ Lett.\  {\bf 116}, no. 13, 132301 (2016).

\bibitem{Tian:2017qwk}
  F.~Tian, C.~Gong and B.~Q.~Ma,
  Nucl.\ Phys.\ A {\bf 968}, 379 (2017)
  doi:10.1016/j.nuclphysa.2017.09.006
  [arXiv:1706.01411 [hep-ph]].



\bibitem{GolecBiernat:1998js}
  K.~J.~Golec-Biernat and M.~W\"usthoff,
  Phys.\ Rev.\ D {\bf 59}, 014017 (1998)
  [hep-ph/9807513].


\bibitem{Mueller:2012uf}
  A.~H.~Mueller, B.~W.~Xiao and F.~Yuan,
  Phys.\ Rev.\ Lett.\  {\bf 110}, no. 8, 082301 (2013)
  [arXiv:1210.5792 [hep-ph]];
  Phys.\ Rev.\ D {\bf 88}, no. 11, 114010 (2013)
  [arXiv:1308.2993 [hep-ph]].


\bibitem{Balitsky:2015qba}
  I.~Balitsky and A.~Tarasov,
  JHEP {\bf 1510}, 017 (2015)
  doi:10.1007/JHEP10(2015)017
  [arXiv:1505.02151 [hep-ph]];
  JHEP {\bf 1606}, 164 (2016)
  doi:10.1007/JHEP06(2016)164
  [arXiv:1603.06548 [hep-ph]].


\bibitem{Marzani:2015oyb}
  S.~Marzani,
  Phys.\ Rev.\ D {\bf 93}, no. 5, 054047 (2016)
  [arXiv:1511.06039 [hep-ph]].

\bibitem{Zhou:2016tfe}
  J.~Zhou,
  JHEP {\bf 1606}, 151 (2016)
  doi:10.1007/JHEP06(2016)151
  [arXiv:1603.07426 [hep-ph]].

\bibitem{Xiao:2017yya}
  B.~W.~Xiao, F.~Yuan and J.~Zhou,
  Nucl.\ Phys.\ B {\bf 921}, 104 (2017)
  doi:10.1016/j.nuclphysb.2017.05.012
  [arXiv:1703.06163 [hep-ph]].


\bibitem{Ji:2004xq}
  X.~d.~Ji, J.~P.~Ma and F.~Yuan,
  Phys.\ Lett.\ B {\bf 597}, 299 (2004)
  doi:10.1016/j.physletb.2004.07.026
  [hep-ph/0405085].


\bibitem{Aybat:2011zv}
  S.~M.~Aybat and T.~C.~Rogers,
  Phys.\ Rev.\ D {\bf 83}, 114042 (2011)
  [arXiv:1101.5057 [hep-ph]].

\bibitem{Marquet:2009ca}
  C.~Marquet, B.~W.~Xiao and F.~Yuan,
  Phys.\ Lett.\ B {\bf 682}, 207 (2009)
  doi:10.1016/j.physletb.2009.10.099
  [arXiv:0906.1454 [hep-ph]].












\end{thebibliography}
\end {document}